# Characterizing Semantic Ambiguity of the Materials Science Ontologies

*Scott McClellan*
Metadata Research Center, Drexel University, U.S.A.
*Yuan An*
Metadata Research Center, Drexel University, U.S.A.
*Xintong Zhao*
Metadata Research Center, Drexel University, U.S.A.
*Xia Lin*
Metadata Research Center, Drexel University, U.S.A.
*Jane Greenberg*
Metadata Research Center, Drexel University, U.S.A.

## *Abstract*

Growth in computational materials science and initiatives such as the Materials Genome Initiative (MGI) and the European Materials Modelling Council (EMMC) has motivated the development and application of ontologies. A key factor has been increased adoption of the FAIR principles, making research data findable, accessible, interoperable, and reusable (Wilkinson et al. 2016). This paper characterizes semantic interoperability among a subset of materials science ontologies in the MatPortal repository. Background context covers semantic interoperability, ontological commitment, and the materials science ontology landscape. The research focused on MatPortal's two interoperability protocols: LOOM *term matching* and *URI matching*. Results report the degree of overlap and demonstrate the different types of ambiguity among ontologies. The discussion considers implications for FAIR and AI, and the conclusion highlight key findings and next steps.

## *1.0 Introduction*

Ontologies, as structured knowledge systems, support information organization, discovery, and retrieval both within and among communities. Moreover, they describe fields of knowledge and encode both conceptual and *real-world relationships* (Gruber 1993). Biagetti (2020) describes ontologies as follows: "ontologies are considered either a conceptual, semantic-level framework, or a concrete artifact provided for a specific purpose," and this dual nature raises problems of "terminological ambiguity." Interconnected here is the relationship among ontologies as reflected in overlapping terminology, specifically terms representing *semantic interoperabilty*. This topic gained greater significance, as researchers seek to ontologies to support FAIR data and AI operations (Voigt and Kalidindi

2021).

Semantic interoperability, specifically term overlap, has been studied extensively in domains such as biology, biomedicine, and agriculture, within the context of ontology repositories, like BioPortal and AgroPortal (Kamdar et al. 2017, Laadhar et al. 2020). Less attention has been given to this topic in the materials science space, likely due to more recent interest in this discipline compared to biology, biomedicine, and other domains The limited research here, along with the global adoption of FAIR and increased interest and growth in materials science ontologies, point to an opportunity to study semantic interoperability across these domain ontologies. The research presented in this paper considers this opportunity and examines MatPortortal's two semantic interoperability protocols (LOOM term overlap and URI matching), and explores some of the complications and issues they present and their implications for FAIR and AI. The goal of the research is to gain insight into semantic ambiguity among materials science ontologies through term overlapping analysis. We pursued the following research objectives:

- Measured the degree of term overlap for a sample of MS ontologies from MatPortal
- Lexical OWL Ontology Matcher (LOOM): Algorithm which matches two terms based on lexical similarity
- URI: Algorithm which matches term URIs between two ontologies
- Examined the types of semantic ambiguity across a subsection of MatPortal ontologies.

To our knowledge, this is one of the first studies characterizing semantic overlap of materials science ontologies in order to understand semantic ambiguity.

This paper is organized in the following manner. The background section discusses semantic interoperability and provides an overview of the materials science ontology landscape. Next, we present our research objectives and procedures, followed by the results. Explored in the discussion are research implications and, finally, highlighted in the conclusion are key results and next steps.

## *2.0 Term Overlap and Semantic Ambiguity*

Term overlap occurs when the symbolic representation overlaps across two or more textual resources, whereas semantic interoperability involves overlap or equivalence of meaning and seeks to address "inconsistencies in terminology and meanings" (Zeng 2018). The symbolic representation with term overlap generally follows *letter* by *letter*, and *word* by *word*, and may rely upon various stemming or word order algorithms to account for minor linguistic variations. There are two basic types of term overlap renderings.

First, term overlap can reveal true semantic interoperability. In this case the linguistic context confirms the same meaning of the overlapping terms. Second, term overlap can present distinct linguistic contexts, illustrating homonymy. A classic example of homonymy is demonstrated with the term *plant*, which may represent flora in the biological sense, or an industrial construction, where items are manufactured. This latter instance represents the contextual problems present in semantic ambiguity which interfere with semantic interoperability. Research pursuing term overlap presents one way to examine an aspect semantic interoperability, even if term overlap does not always represent a valid semantic overlap.

### *3.0 The Materials Science Ontological Landscape*

Materials science (MS) functions as an umbrella term for a wide range of sub-domains which vary in scale from the atomic to the architectural. In addition to this a variety of disciplines structure and influence materials science, ranging from chemistry and physics to design and engineering (Ashby et al. 2010). Due to this, the materials science landscape reflects a wide variation of ontological systems. These systems are stratified according to the level of abstraction or detail they present. This stratification of ontologies has led to a landscape of materials science ontologies which often describe a limited subdomain area within the context of a specific project. For example, the early Plinius ontology for ceramics evinces this aspect: "The ontology covers the domain of ceramic materials and is developed specifically for the Plinius project (van der Vet et al. 1994, p. 2)." Similarly, the more recent Laser Powder Bed Fusion Ontology (LPBFO) was developed by Fraunhofer to streamline industrial processes. Recent efforts such as the European Union's Horizon 2020 project and the United States Materials Genome Initiative (MGI) have bolstered research into knowledge graphs based on ontologies for data-driven materials science research.

The field of materials science ontologies is a complex space of knowledge representations which includes not only ontologies but also thesauri and knowledge graphs. Overlap is crucial to ensuring the preservation of semantics across knowledge representations. Methods of semantic disambiguation for similar and identical terms have been studied in biology and medicine which may similarly be applicable to MS ontologies (Kamdar et al. 2017). Researchers explored the deployment of the explicit reference (*xref*) syntax in BioPortal which allowed for cross-referencing terms without using the OWL *sameas* syntax, and the results showed an unintended and often ambiguous usage of xref, creating connections which were outside the scope of the intended syntax (Kamdar et al. 2017). This research offered the initial motivation for this study; however, MatPortal does not use the *xref* syntax, but it does utilize automatic ontology term matching algorithms. This study seeks to identify the types of semantic ambiguity that occur due to term overlap in the MatPortal repository that arise from the results yielded from automatic matching.

The MatPortal repository is a curated set of materials science ontologies and vocabularies that are accessible through an application platform based on BioPortal. The ontologies hosted on MatPortal represent different degrees of knowledge including abstract domain ontologies, mid-level ontologies, and area-specific ontologies. This arrangement connects a wide range of representations of the materials science space, and it allows users to explore these structures using matching tools based on two protocols: 1) semantic matching, which starts with letter-by-letter and word-by-word and results in a true semantic equivalent, and 2.) URI matching, which matches terms based on their URIs.

## *4.0 Research Methods and Procedures*

Two term overlapping analysis methods, *automatic mapping* followed by *crosswalk analysis*, were used in this research. The automatic mapping approach leverages standard ontology encoding, allowing us to generate lists of overlapping terms between two selected ontologies. Mappings either matched distinct URIs or term-level semantics, the latter utilized the Lexical OWL Ontology Matcher (LOOM) algorithm (Ghazvinian et al. 2009). LOOM creates mappings using the following procedure:

In order to identify the correspondences, LOOM compares preferred names and synonyms of the concepts in both ontologies. It identifies two concepts from different ontologies as similar, if and only if their preferred names or synonyms are equivalent based on a modified string-comparison function (Ghazvinian et al. 2009, p. 199).

The crosswalk analysis method was supported by human verification of the terminological matches and analyzed some of the semantic issues involved in utilizing such a procedure. The basic approach to systematized mapping work was facilitated by MatPortal's infrastructure and use of OWL, RDF/XML and standards articulated by researchers to support interoperability and term mapping across thesauri and other vocabularies (Clarke and Zeng 2012, Roe and Thomas 2013).

Data was collected using the MatPortal ontology mapping tool which matches terms between two selected ontologies using the following two methods:

1. URI matching: This method compares the underlying URI for terms in an ontology, returning a positive SAME_URI if the values were exact.
2. LOOM matching: This method utilizes the LOOM algorithm which compares terms at several semantic levels within an ontology including any synonyms embedded within the OWL code (Ghazvinian et al. 2009).

Matching results from these two methods were downloaded, and then coded into a series of spreadsheets to examine the features present. The research focused on a convenient sample of five ontologies, selected, based largely upon their overall representation of LOOM and SAME_URI overlap. Table 1 shows the the number of overlaps between the terms of the ontology named and all other ontologies located in MatPortal at the time the study was performed. The subset of ontologies chosen show a diversity of overlapping terms but between themselves only account for approximately 15% of all overlapping terms among all MatPortal ontologies. Moreover, the choices attempt to mitigate the relatively insular set of origins for ontologies in MatPortal several produced by Fraunhofer and or the Materials Open Lab (Matolab) project, which is described as a "venture between Fraunhofer Alliance MATERIALS and Bundesanstalt für Materialforschung und -prüfung (BAM) ("PP20 Mat-o-Lab—Materials-open-Laboratory")." In addition to this, the ontologies chosen attempt to represent different subdomains of MS as well as varying levels of abstraction. The five ontologies selected for analysis included in this study are presented in the following key:

Key to ontologies:

- MaterialsMine (MM)
- Materials Science and Engineering Ontology (MSEO)–
- BWMD Domain Ontology (BWMD_DOM)
- Laser Power Bed Fusion Ontology (LPBFO)
- Matolab Tensile Test Ontology (MOL_TENSILE)

| Ontology Name | Total Classes | URI Matches | LOOM Matches |
|---|---|---|---|
| MSEO | 1657 | 1563 | 767 |
| BWMD_DOM | 772 | 1075 | 543 |
| LPBFO | 509 | 1074 | 447 |
| MOL_TENSILE | 372 | 869 | 439 |
| MM | 2052 | 4 | 998 |

Table 1. Values of MatPortal ontology overlap using both URI and LOOM matching methods. These values show matches with all other ontologies in MatPortal; matches among the sample are shown in the Results section.

### *5.0 Results*

The results reported here cover aggregate counts for matching, an overview and the specific count for URI matching and the LOOM matching algorithms. Each of the semantic and URI matching techniques produced distinct results which point toward patterns of overlap which are both beneficial and detrimental to varying degrees discussed below. The values and syntactical samples used for this study derive from the RDF/XML implementations of the ontologies unless otherwise noted.

### *5.1 URI Matching*

| | BWMD_DOM (772) | LPBFO (509) | MSEO (1657) | MM (2052) | MOL_TENSILE (372) |
|---|---|---|---|---|---|
| BWMD_DOM (772) | | 347 | 35 | 0 | 347 |
| LPBFO (509) | 347 | | 35 | 0 | 347 |
| MSEO (1657) | 35 | 35 | | 0 | 35 |
| MM (2052) | 0 | 0 | 0 | | 0 |
| MOL_TENSILE (372) | 347 | 347 | 35 | 0 | |
| Total | 729 | 729 | 105 | 0 | 729 |

Table 2. URI Matching Data Among Sample Ontologies. Values in parentheses show the number of classes for each ontology.

Table 2 displays the number of terms identified by the URI matching algorithm which match between pairs of ontologies; parenthetical values show the total number of terms for each ontology. URI matching exposed a division in the data with two distinct subsets arising. The first of these subsets, MSEO and MM, showed few to no overlaps among the compared ontologies. MM displayed no URI matches with the sample subset but has the most classes, as can be seen in Table 2 above. These statistics reflect the lack of URI matches for MM displayed in Table 1 above. MSEO returned 35 URI matches each with BWMD_DOM, LPBFO, and MOL_TENSILE despite having 1657 terms of its own. Moreover, overlap viewed as a percentage of ontology terms varies widely with MOL_TENSILE sharing 93% of its terms with both BWMD_DOM and LPBFO. The more plentiful URI matches might occur due to either a common source as those between BWMD-DOM, LPBFO, MSEO, and MOL_TENSILE, which share a common point origin at Fraunhofer as noted or from the use of a common abstract framework, in this case BFO. Moreover, MOL_TENSILE and LPBFO both import the BWMD_MID ontology whose terms overlap.

Focusing on connections that occur at the individual term level exposes some complexity in their relationships. In the syntax from the OWL implementation, Example 1, where the term “AdditiveManufacturingMachine” is grafted to the class “EquipmentSet,” we can see that LPBFO uses BWMD_MID as a parent ontology. In this case the connected terms form throught the interaction of the *comment* and *isDefinedBy* properties which conceptually situate the term “AdditiveManufacturingMachine” in the external reality which the ontology seeks to represent.

```
<owl:Class
rdf:about="https://www.emi.fraunhofer.de/ontologies/LPBFO#LPBFO_0
0002">
<rdfs:subClassOf
rdf:resource="https://www.materials.fraunhofer.de/ontologies/BWMD_on
tology/mid#BWMD_00170"/>
<rdfs:comment xml:lang="en">Section of the additive manufacturing
```

```
system, including hardware and machine control software, required
commissioning software and peripheral accessories that are necessary to
complete a construction cycle for producing parts.</rdfs:comment>
<rdfs:isDefinedBy xml:lang="en"> ISO/ASTM
52900</rdfs:isDefinedBy>
<rdfs:label xml:lang="en">AMMachine</rdfs:label>
<rdfs:label xml:lang="en">AdditiveManufacturingMachine</rdfs:label>
</owl:Class>
```

Example 1. ‘Additive Manufacturing Machine’ term from LPBFO.

This usage broadens the representative capacity of the ontology by providing an intersecting set of shared terms, the mid-level ontology in this case, which permits interoperability between the highly localized vocabulary of LPBFO and the more general top-level ontology such as BFO (Huschka 2020).

The use of OWL *import* statements complicates the question of URI matching and is reflected in several of the results (Table 2). The cluster of overlapping terms shared by BWMD_DOM, LPBFO, and MOL_TENSILE are primarily the result of this syntax. This issue is addressed further below in the Discussion section.

Among the most common ontologies that appear in the SAME_URI matches is the Basic Formal Ontology (BFO), which forms a backbone for many ontological structures. Arp et al. (2015) describes it as “an upper-level ontology developed to support integration of data obtained through scientific research.” The Fraunhofer associated ontologies, BWMD-DOM, LPBFO, MOL_TENSILE and MSEO, employ the most recent version of BFO. However, there is an earlier version of BFO which utilizes a different namespace and whose semantics differ. While the earlier version is convertible to the more recent one, the two are not necessarily compatible. This points to a secondary problem where changes to externally maintained ontologies can lead to future semantic ambiguity, reducing interoperability between ontologies, especially when relying on the OWL *import* property.

### *5.2 LOOM (Terminological) Matching*

| | BWMD_DOM (772) | LPBFO (509) | MSEO (1657) | MM (2052) | MOL_TENSILE (372) |
|---|---|---|---|---|---|
| BWMD_DOM (772) | | 17 | 127 | 71 | 6 |
| LPBFO (509) | 17 | | 126 | 63 | 0 |
| MSEO (1657) | 127 | 126 | | 112 | 126 |
| MM (2052) | 71 | 63 | 112 | | 66 |
| MOL_TENSILE (372) | 6 | 0 | 126 | 66 | |
| Total | 221 | 206 | 491 | 312 | 198 |

Table 3. LOOM Matching Data Among Sample Ontologies. Values in parentheses show the number of classes for each ontology.

Table 3 displays the number of terms identified by the LOOM algorithm which match between pairs of ontologies; parenthetical values show the total number of terms for each ontology. The LOOM algorithm returned matches between almost all ontologies in the sample. MM and MSEO displayed the greatest degrees of overlap between the sample ontologies. The remaining three ontologies in the sample showed high degrees of overlap between each other and also with the larger MatPortal repository. Only one pair of ontologies in the sample which show no LOOM matches are MOL_TENSILE and LPBFO. This lack of overlap might be due to the specificity of domains which each ontology covers, or else offset by high levels of URI matches as seen in Table 2 above. MOL_TENSILE and LPBFO both import large numbers of their terms directly from the BWMD ontologies, probably reducing the possible pool of terminological matches. Between individual ontologies, semantic overlap occurs in a somewhat varied pattern with MM and MSEO showing the most consistent pattern of matches with the other ontologies in the sample, as can be seen above in Table 3. The overlap for MM and MSEO with other ontologies could be an effect of their relative size, 2052 and 1657 classes respectively.

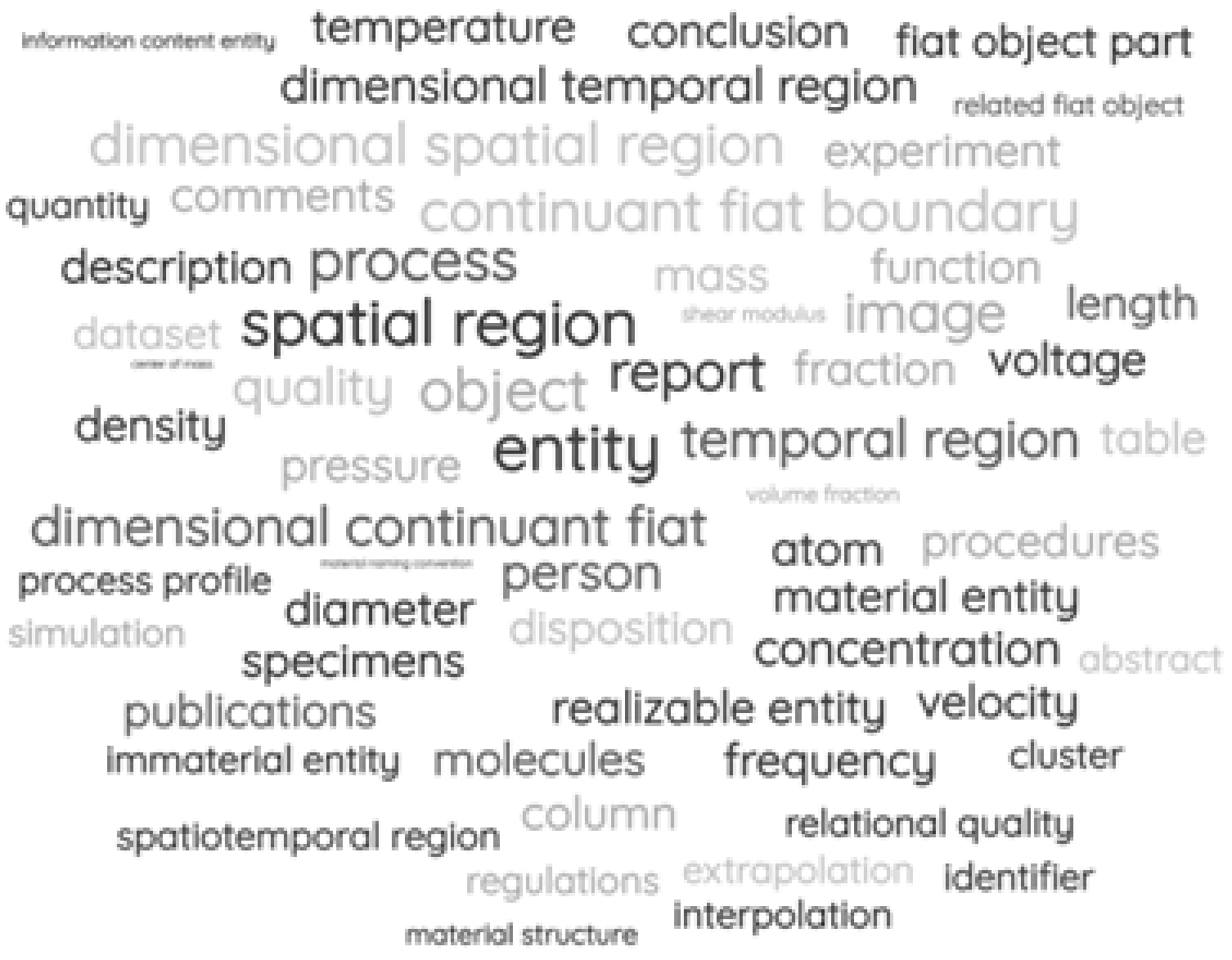

Figure 1. Predominance of abstract terms using LOOM algorithm

At the term level, corresponding elements display several characteristics which require attention due to the questions about semantic ambiguity that they raise. The majority of term overlap tends to occur at more abstract or generic levels rather than at what might be considered the granular, area-specific level. The word cloud in Figure 1 above offers an overview of the terminology, with very few area-specific terms represented and a predominance of abstractions or generic concepts. Some of this overlap occurs at the highly generic level such as terms like 'person' or 'software script' which explicitly provide categories to differentiate commonly occurring instances, and both of these terms occur across all of the ontologies in the sample. In addition, the basis of comparison determines the match which occurs; MM presents a LOOM match for the term 'disposition' with all members of the sample. This aspect is complicated by the fact that the MM term is itself imported from the Semanticscience Integrated Ontology. In addition, matching terms such as 'voltage' or even 'transmission electron microscopy' display specific, if somewhat generic, points of similarity.

Looking at the granular case of the term 'agent' as expressed in the MM and MSEO ontologies some distinct issues with term matching become apparent. MM utilizes three different representations, each of which is slightly different: 1. FOAF, 2. PROV, 3. DCTERMS. Each of these defines 'agent' slightly differently:

- FOAF (http://xmlns.com/foaf/0.1/Agent): "The Agent class is the class of agents; things that do stuff. A well known sub-class is Person, representing people. Other kinds of agents include Organization and Group."
- PROV (http://www.w3.org/ns/prov#Agent): "An agent is something that bears some form of responsibility for an activity taking place, for the existence of an entity, or for another agent's activity."
- DCTERMS (http://purl.org/dc/terms/Agent): "A resource that acts or has the power to act."

MSEO imports its use of agent from the Common Core Ontologies (CCO), which describes it as, "The class AGENT comprises both individual agents (PERSON) and coordinated groups of individuals (ORGANIZATION)" (CUBRC 2021). Though there is overlap at the level of meaning in each definition, there are slight differences between how each is defined, which leads to possible ambiguity at the semantic level.

.

*6.0 Discussion*

The results presented above give insight into some ambiguities in term overlap among a sample of ontologies in MatPortal. These results may aid materials science researchers seeking to leverage ontologies to support the FAIR principles (Wilkinson et al. 2016), data-driven research (Moreno Torres 2021), and even AI (Aggour et al. 2022, Voigt and Kalidindi 2021). For data-driven environments, categorizing large volumes of data and outputs accurately assists in deeloping meaningful results. For AI applications, ontologies and terminologies provide a semantic and logical backbone for knowledge graphs from which meaning derives. The results specifically demonstrate s several problems that arise from the term overlaps. For example, URI matching results provide isomorphic mappings between terms, limiting ambiguity. This can offer a high degree of interoperability, however, many of these terms occur in ontologies produced by a small number of institutions, such as BWMD and Fraunhofer, which share similar research projects. This is not to say that such ontologies are not representative of the research landscape, but they are limiting. The overlapping terms tend to be the result of the OWL *import* function which includes all terms whether representative of the domain or not from the imported ontology, and this leads to an excess of terms that are not relevant. Moreover, importing an ontology in OWL is transitive leading to a case where terminology and logic could lead to semantic ambiguity as was the case with the analysis of the term ‘agent’ where definitions within and among ontologies can vary by degrees (Antoniou et al 2012).

While LOOM matches decrease when URI matches increase in the sample, the relationship is not strictly inverse. This may indicate that the two types of term overlap account for different features of the ontologies. For example, terms identified by the LOOM algorithm seem to display greater ambiguity as they can occur multiple times between ontologies, referring to similar definitions located at different URIs. Overall, these results give insight into the ambiguities that researchers as well as systems face when trying to leverage ontologies for data interoperability. Further research in this area and gaining a deeper understanding of these relationships will allow for more precise semantic interoperability by better defining the lexical connections between ontological systems.

## *7.0 Conclusion*

This paper characterized semantic ambiguity as it relates to interoperability across a sample of materials science ontologies, specifically through term overlap. The research focused on a subset of five MatPortal materials science ontologies. Two separate automatic indexing algorithms were employed: one which assessed terms based on term similarity and another which matched terms based on identical URIs. These terms were then analyzed using crosswalk analysis to see possible types of ambiguity which could arise from term overlap.

The analyses found that both term and URI matching revealed different types of ambiguity between ontologies. In URI matching, high levels of overlap were often the result of ontologies importing related area ontologies or else upper- or domain-level ontologies. Importing an outside ontology artificially inflates the quantifiable overlap, and, in the case of this study, is highly dependent on the sample as can be seen with discussion of MSEO and CCO above. Term matching presented a different concern related to the meanings which underpinned overlapping terms; differences in meanings, especially for terms which occur multiple times where each could reference a separate resource. This structure entails several ontological commitments creating some confusion regarding what is meant by a term, such as the case with 'agent' above.

Although this research employs a basic set of methods applied to a limited sample, the results lay groundwork for a number of next steps. One key direction is to explore the complexity of importing external ontologies and their impacts on URI matching in MatPortal and the broader materials science ontology interoperability is one avenue to examine. Another direction is to investigate term matching and how it affects meaning as well as ontological commitments is also of great importance in understanding interoperability.

In conclusion, the work reported in this paper presents preliminary look at the problem of term overlap in a limited setting which points toward the need for a broader engagement of the subject of interoperability of terminology in the MS semantic space. Furthermore, this paper contributes a methodological approach for future ressearch in this area. Finally, the ground work and the research direction covered in this paper has particular importance, as a proper aligned set of representations both within the ontology and real-world spaces of materials science is imperative for effective use of ontologies in AI and other machine-driven research. This research provides a preliminary look at the problem of term overlap in a limited setting which points toward the need for a broader engagement of the subject of interoperability of terminology in the MS semantic space.

## *8.0 Acknowledgments*

The authors acknowledge funding for this paper was provided by the National Science Foundation, grant OAC 2118201, the Institute for Data-driven Dynamical Design.